\documentclass[a4paper,12pt]{article}
\usepackage[english]{babel}
\usepackage{natbib, url}
\usepackage{float,graphicx,subfigure}

\newcommand{\Fig}[1]{Fig. \ref{#1}}
\newcommand{\Sec}[1]{Sect. \ref{#1}}

\begin{document}

\bibpunct{(}{)}{,}{}{}{,}
\bibliographystyle{dcu}

\title{\bf Contextual emergence of intentionality}
\author{
 Peter beim Graben \\
 Department of German Studies and Linguistics, and \\
 Berlin School of Mind and Brain \\
 Humboldt-Universit\"at zu Berlin
 }
\date{\today}

\maketitle

\begin{abstract}
By means of an intriguing physical example, magnetic surface swimmers, that can be described in terms of Dennett's intentional stance, I reconstruct a hierarchy of necessary and sufficient conditions for the applicability of the intentional strategy. It turns out that the different levels of the intentional hierarchy are contextually emergent from their respective subjacent levels by imposing stability constraints upon them. At the lowest level of the hierarchy, phenomenal physical laws emerge for the coarse-grained description of open, nonlinear, and dissipative nonequilibrium systems in critical states. One level higher, dynamic patterns, such as, e.g., magnetic surface swimmers, are contextually emergent as they are invariant under certain symmetry operations. Again one level up, these patterns behave apparently rational by selecting optimal pathways for the dissipation of energy that is delivered by external gradients. This is in accordance with the restated Second Law of thermodynamics as a stability criterion. At the highest level, true believers are intentional systems that are stable under exchanging their observation conditions.
\end{abstract}

%--------------------------------------- Section -------------------------------------------------------------

\section{Introduction}
\label{intro}

\begin{figure}[H]
  \centering
  \includegraphics[width=\textwidth]{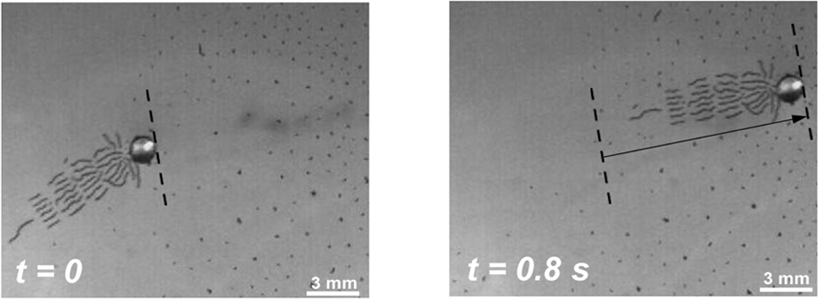}
  \caption{\label{fig:surfswim} Magnetic surface swimmer. Reprinted from \citet[Fig. 1(d)]{SnezhkoBelkinEA09} [Snezhko, A., Belkin, M., Aranson, I. S. \& Kwok, W.-K. \emph{Physical Review Letters}, 102, 118103, 2009] with permission. Copyright (2009) by the American Physical Society.
  }
\end{figure}

Once upon a time, physicists A. Snezhko and I. S. Aranson from Argonne National Laboratory in Illinois conducted an experiment on the coupling of the electromagnetic field to hydrodynamics \citep{SnezhkoAransonKwok05}. They suspended nickel microspheres over the surface of water in a beaker and applied an alternating magnetic field perpendicular to the liquid-air interface. Following the magnetic field, the spheres, supported by the fluid's surface tension, began to rotate thereby causing vortices that changed the spheres' local environment. As an intriguing consequence, processes of self-organization and self-assembly took place: spheres aligned up to chains; chains arranged in parallel to form segments and segments created large-scale patterns reminiscent to worms or snakes \citep{SnezhkoAransonKwok06, BelkinSnezhkoEA07}. Yet these structures were not only static patterns, they dynamically generated streams which were expelled by the two tails of the snake. Hence, a snake stalled as a highly efficient water pump surrounded by a quadrupolar velocity field \citep{BelkinSnezhkoEA07}. Tuning the driving frequency of the magnetic field, the system exhibited a dynamic instability when the pump of one tail got stronger than the pump of the other tail. As a consequence of this symmetry breaking, the snake started to move around \citep{SnezhkoBelkinEA09}. Moreover, this symmetry breaking could has been caused also by putting a plastic bead into the container. After some time of drifting around, the bead got incorporated into a snake, thus becoming its distinguished head by interrupting the water stream at this site \citep{SnezhkoBelkinEA09}. The resulting complex object, shown in \Fig{fig:surfswim}, erratically swam around, exploring its environment and snatching isolated nickel spheres that were also incorporated into the structure. Thereby, the snake grew and accelerated further, becoming an even more awesome predator.

Snezhko and Aranson's discovery of these co-called \emph{magnetic surface swimmers} \citep{SnezhkoBelkinEA09} has been brought to the public by the \emph{Wired Science} blog of Madrigal:\footnote{
    \url{http://www.wired.com/wiredscience/2009/03/snakes/} where also a collection of movies and further details is available (see also Snezhko and Aranson's supplement \url{http://mti.msd.anl.gov/highlights/snakes/}).
}
\begin{quote}
    These chains of metal particles look so much like real, living animals, it is hard not to think of them as alive. \dots As it starts heading for other chains of particles in an unpredictable and eccentric way, it's nearly impossible not to anthropomorphize the structure. It just acts too much like life. The damn thing practically has \dots personality. \dots `It also has a very bad temper,' Aranson jokes, noting that this creature, this figment of nature, appears to `hunt' the other particles. Indeed it does. As you can see in the video, the metallic monster, technically known as a `surface swimmer,' acts hungry. As it snatches more particles, it swims faster and faster.
\end{quote}

Indeed, looking at the videos at \emph{Wired Science}, it is tempting to regard the snakes as hungry predators, desiring to catch as many nickel spheres as possible and believing that they have to go stalking for them as they eagerly demand food for survival. In other words, on the one hand, it is almost inevitable to adopt the \emph{intentional stance} \citep{Dennett78, Dennett89} in order to describe and predict the behavior of these structures by ascribing beliefs and desires to them. On the other hand, the physics of magnetic surface swimmers is completely understood in terms of the laws of Newtonian mechanics, Maxwell's electrodynamics and the phenomenal Navier-Stokes equation governing hydrodynamics \citep{BelkinGlatzEA10}. But this puts physicists such as Snezhko and Aranson to the very same place of the `cleverer observer', \citet[pp. 23]{Dennett89b} discusses in his famous Martians example \citep[pp. 25]{Dennett89b} in order to address the alleged observer-dependence of intentionality and hence of consciousness in general.

Therefore, the magnetic surface swimmer example illustrates firstly that being a `cleverer observer' does not prevent one from employing the intentional strategy. But secondly, as I argue in this article, it also shows that intentionality cannot be reduced to Laplacean physics because physical science actually works very differently: The phenomenal laws of, e.g., thermodynamics or hydrodynamics, are not completely reducible to a fundamental level of, say, elementary particle physics (that is metaphorically often conceived as classical point mechanics in the physicalist literature). The reason for this is that fundamental levels of description only provide necessary but not sufficient conditions for a description at a higher level. Sufficient conditions have to be provided by contingent contexts which implement stability conditions for the system's dynamics at the lower level. Such a non-reductive relation between different levels for the description of systems has been introduced by \citet{BishopAtmanspacher06} and  \citet{AtmanspacherBishop07} as \emph{contextual emergence}. In particular, \citet{Bishop08, Bishop12} demonstrates how the nonlinear and dissipative properties of fluid dynamics emerge upon purposefully and rationally chosen contexts.

The article is structured in the following way: In \Sec{intsys} I briefly review Dennett's operationalistic definition of intentional systems, while \Sec{contemer} delivers an overview about the theory of contextual emergence. In \Sec{ce} I reconstruct the contextual emergence of intentionality in four steps, comprising a hierarchy of necessary and sufficient conditions: Subsection \ref{ce1} discusses the emergence of those nonlinear laws that govern the self-organization of dissipative nonequilibrium structures such as magnetic surface swimmers. Subsection \ref{ce2} then describes the emergence of these particular structures by means of self-assembling patterns. Subsection \ref{ce3} applies the concepts of contextual emergence to Dennett's definition of \emph{intentional systems}. Subsection \ref{ce4} concludes with the emergence of what  \citet[p. 27]{Dennett89b} calls `true believers'. In \Sec{disc} I resume the discussion concluding that \emph{true believers are real observers}.

%--------------------------------------- Section -------------------------------------------------------------

\section{Intentional Systems and True Believers}
\label{intsys}

Dennett's operationalistic definition of intentionality \citep{Dennett78a, Dennett89b} neglects the questions whether conscious beings `\emph{really} have beliefs and desires' \citep[p. 22]{Dennett89b} and whether they are localized somewhere `in the believer's head' \cite[p. 14]{Dennett89b}.\footnote{
    Or whether they are somehow neurally correlated \citep{Chalmers00}).
}
Dennett's intentional strategy is one of three possible strategies that can be employed \emph{by an observer} who is asked to describe and predict some system's behavior: `\dots a particular thing is an intentional system only in relation to the strategies of someone who is trying to explain and predict its behavior' \citep[pp. 3f]{Dennett78a}.

Exemplified by means of a chess computer, \citet[pp. 4ff]{Dennett78a} demonstrates that an observer can firstly adopt the \emph{physical stance} (cf. \citet[p. 16]{Dennett89b}) regarding the system as a physical object comprised from molecules, atoms or elementary particles (or interacting fields, as a physicists could add) and governed by deterministic natural laws. In the ideal case, the observer adopting the physical strategy is a Laplacean demon who is able to apply his complete knowledge of natural laws to his complete knowledge of initial conditions in order to compute any future state of the system under study \citep[p. 23]{Dennett89b}.

The second possibility for an observer is to adopt the \emph{design stance} (\citet[p. 4]{Dennett78a}, \citet[pp. 16f]{Dennett89b}; sometimes referred to as the \emph{functional stance} \citep{Bieri97a}). The chess computer subjected to the design stance is described by an algorithm that computes and weights all possible moves in order to decide for the optimal one because it has actually been designed for that aim \citep[p. 4]{Dennett78a}.

Finally, the observer may adopt the \emph{intentional stance} (\citet[p. 5]{Dennett78a}, \citet[p. 17]{Dennett89b}):

\begin{quote}
Sometimes even the design stance is practically inaccessible, and then there is yet another stance or strategy one can adopt: the intentional stance. Here is how it works: first you decide to treat the object whose behavior is to be predicted as a rational agent; then you figure out what beliefs that agent ought to have, given its place in the world and its purpose. Then you figure out what desires it ought to have, on the same considerations, and finally you predict that this rational agent will act to further its goals in the light of its beliefs. \citep[p. 17]{Dennett89b}
\end{quote}

For the chess computer this means to treat `the machine rather like an intelligent human opponent' who rationally accepts the goal of the game and its rules \citep[p. 5]{Dennett78a}. Then the observer ascribes `to the system the \emph{possession of certain information}' i.e. beliefs and supposes `it to be \emph{directed by certain goals}', i.e. desires \citep[p. 6]{Dennett78a}. In other words:

\begin{quote}
Having doped out these conditions \dots we can proceed at once to ascribe beliefs and desires to the creatures. Their behavior will `manifest' their beliefs by being seen as the actions which, given the creatures' desires, would be appropriate to such beliefs as would be appropriate to the environmental stimulation. Desires, in turn, will be `manifested' in behavior as those appropriate desires (given the needs of the creature) to which the actions of the creature would be appropriate, given the creatures beliefs. The circularity of these interlocking specifications is no accident. Ascriptions of beliefs and desires must be interdependent \dots. \citep[pp. 8f]{Dennett78a}
\end{quote}

This is an interesting point because the intentional stance here resembles the `circular causality' in the self-organization of complex systems as described by synergetics \citep{Haken83a, TschacherHaken07}. I come back to this issue in the following sections.

\citet[pp. 7f]{Dennett78a} concludes:

\begin{quote}
Lingering doubts about whether the chess-playing computer \emph{really} has beliefs and desires are misplaced, for the definition of intentional systems I have given does not say that intentional systems \emph{really} have beliefs and desires, but that one can explain and predict their behavior by \emph{ascribing} beliefs and desires to them. \dots All that has been claimed is that on occasion, a purely physical system can be so complex, and yet so organized, that we find it convenient, explanatory, pragmatically necessary for prediction, to treat it as if it had beliefs and desires and was rational.
\end{quote}

This nicely illustrates Dennett's operationalistic or instrumentalistic attitude. But now another problem arises:

\begin{quote}
The next task would seem to be distinguishing those intentional systems that \emph{really} have beliefs and desires from those we may find it handy to treat \emph{as if} they had beliefs and desires. \dots A better understanding of the phenomenon of belief begins with the observation that even in the worst of these cases, even when we are surest that the strategy works \emph{for the wrong reasons}, it is nevertheless true that it does work, at least a little bit. This is an interesting fact, which distinguishes this class of objects, the class of \emph{intentional systems}, from the class of objects for which the strategy never works. \citep[pp. 22f]{Dennett89b}
\end{quote}

Hence, the definition of an intentional system is that of a system which can be successfully predicted by virtue of the intentional strategy. Moreover: `\emph{all there is} to being a true believer is being a system whose behavior is reliably predictable via the intentional strategy' \citep[pp. 29]{Dennett89b}. However, the intentional strategy may work `for the wrong reasons'. Thus, the last quotation has to be interpreted in the following way: Every true believer is an intentional system, but not every intentional system must be a true believer. That is, the intentional stance only provides a necessary condition for being a true believer, while a sufficient condition must be sought somewhere else.

According to the operationalistic definition of intentionality it would be `intolerable to hold that some artifact or creature or person was a believer
from the point of view of one observer, but not a believer at all from the point of view of another, cleverer observer' \citep[pp. 23f]{Dennett89b}.  \citet[p. 25]{Dennett89b} illustrates this problem by a famous \emph{Gedankenexperiment}:

\begin{quote}
Suppose \dots some beings of vastly superior intelligence --- from Mars, let us say --- were to descend upon us, and suppose that we were to them as simple thermostats are to clever engineers. Suppose, that is, that they did not \emph{need} the intentional stance --- or even the design stance --- to predict our behavior in all its detail. They can be supposed to be Laplacean super-physicists, capable of comprehending the activity on Wall Street, for instance, at the microphysical level. Where we see brokers and buildings and sell orders and bids, they see vast congeries of subatomic particles milling about ---  and they are such good physicists that they can predict days in advance what ink marks will appear each day on the paper tape labeled `Closing Dow Jones Industrial Average.' They can predict the individual behaviors of all the various moving bodies they can observe without ever treating any of them as intentional systems. Would we be right then to say that from \emph{their} point of view we really were not believers at all (any more than a simple thermostat is)? If so, then our status as believers is nothing objective, but rather something in the eye of the beholder --- provided the beholder shares our intellectual limitations.
\end{quote}

\citet[p. 25]{Dennett89b} immediately retorts:

\begin{quote}
Our imagined Martians might be able to predict the future of the human race by Laplacean methods, but if they did not also see us as intentional systems, they would be missing something perfectly objective: the \emph{patterns} in human behavior that are describable from the intentional stance, and only from that stance, and that support generalizations and predictions.
\end{quote}

Stressing the Martians argument even further, \citet[pp. 25f]{Dennett89b} concludes that intentional acts are abstract and therefore not accessible at the fundamental level of the physical stance. Finally, he invents a kind of Turing test \citep{Turing50} where an Earthling disguises himself as a Martian to become treated as a proper intentional system. This eventually reveals for \citet[p. 27]{Dennett89b} `\emph{the unavoidability of the intentional stance with regard to oneself and one's fellow intelligent beings}.'

%--------------------------------------- Section -------------------------------------------------------------

\section{Contextual Emergence}
\label{contemer}

In the previous section I have already elucidated that the intentional stance only supplies a necessary condition for being a true believer while the sufficient conditions have to be somehow reconstructed from the concluding period. Therefore it sounds reasonable to carry out this reconstruction in terms of contextual emergence (\citet{BishopAtmanspacher06, AtmanspacherBishop07}, see also \citet{AtmanspacherGraben09} for an overview).

In general, science often considers different levels for describing one and the same system. Physics, e.g., looks at an ideal gas in thermal equilibrium as a huge collection of essentially non-interacting microscopic particles that freely move around in a container colliding with each other and bouncing the container's boundaries in a deeply elastic manner. On a macroscopic level, by contrast, an ideal gas in thermal equilibrium is described by the phenomenal quantities pressure, volume and temperature that obey the ideal gas law \citep{PitaevskiiLifshitz81}. As another example, the microscopic state of a neural network is given through the activity of a large number of individual neurons. On the other hand, macroscopic patterns of distributed activity lead to neural macrostates serving as cognitive representations in connectionist modeling \citep{Amari74, GrabenBarrettAtmanspacher09}.

These examples raise the question of the logical connection between different levels of description. A dependency that is very prominent in the philosophical literature is \emph{strong reduction} when a lower level provides necessary and sufficient conditions for a higher level. Another important example, \emph{supervenience}, refers to the case where a lower level only provides sufficient but not necessary conditions for a higher level which means that a higher level description could be multiple realized at the lower level. Less interesting is the possibility that there are neither necessary nor sufficient conditions between two levels, a case that has been referred to as \emph{radical emergence}. The last option offers necessary but not sufficient conditions for the higher level at the lower level. If these are provided by a contingent context that implements stability conditions at the lower level for the description at the higher level, \citet{BishopAtmanspacher06} call this \emph{contextual emergence}.

The importance of stability conditions resulted from a detailed investigation of some peculiarities of quantum physics \citep{Primas90a}. Some of the most pertinent philosophical problems of quantum theory, notably the famous \emph{measuring problem}, but also the emergence of classical quantities such as temperature in quantum statistical mechanics, were not soluble in von Neumann's canonical codification of quantum mechanics \citep{Neumann55}. However, these problems became tractable in the more powerful formulation of algebraic quantum theory \citep{Hepp72, Takesaki70, Primas00}.

Considering the paradigmatic example of thermodynamics again, one firstly has to define the two levels of description. Clearly, the lower level comprises the movements and collisions of a huge number of individual particles, described by a point in high-dimensional phase space that is spanned by the particles' positions and momenta in the classical picture of point mechanics. In the same picture, the higher level is then prescribed by the space of probability distribution functions over phase space, called statistical states. These are the referents of classical statistical mechanics.\footnote{
    Likewise, statistical states in quantum statistical mechanics are given as density matrices.
}
However, this statistical state space of probability distribution functions is much too large to bear physically meaningful interpretations.\footnote{
    In this respect, \citet[p. 241]{Primas90a} talks about a mathematical `monster, which contains myriads of unphysical states'.
}
Therefore, the space of statistical states has to be substantially restricted to become operationally and epistemically well-defined. This is achieved by the  observer who chooses a contingent context from a particular point of view \citep{Graben2011}, defining the \emph{relevant properties} of the problem under study.\footnote{
    Cf. also \citet[p. 16]{Dennett89b}: where `one ignores the actual (possibly messy) details' for the design stance.
}

Deliberately choosing a contingent context, introduces a coarse-graining where singularities at the lower level are smoothed out at the higher level by changing the problem's topology.\footnote{
    Similarly such smoothing could be gained by asymptotic expansions, e.g. for separating timescales \citep{Primas98, BishopAtmanspacher06, AtmanspacherGraben07}.
}
However, not every arbitrarily selected context entails emergent properties. This requires the implementation of stability conditions at the lower level as a sufficient condition for the higher level description. For the example of thermodynamics, stability refers to the concept of \emph{thermal equilibrium} which is alien to statistical mechanics. In physics textbooks, the notion of thermal equilibrium is usually introduced by the Zeroth Law, presenting an equivalence relation between physical systems: An equilibrium system $A$ is in equilibrium with itself (reflexivity). When system $A$ is in equilibrium with system $B$, then $B$ is in equilibrium with $A$ also (symmetry). When $A$ is in equilibrium with $B$ and $B$ with $C$ then $A$ is in equilibrium with $C$ (transitivity). By virtue of this relation, all systems that are in thermal equilibrium with each other, become operationally indistinguishable, thus forming an equivalence class, or likewise, the pattern at a higher level of description.

At the lower level of description, thermal equilibrium states can be identified with particular statistical states, called Kubo-Martin-Schwinger (KMS) states \citep{Primas98, BishopAtmanspacher06}. These are characterized by three stability conditions: (1) temporal stability, (2) structural stability, and (3) `forgetfulness' \citep{HaagKastlerTrych-Pohlmeyer74, AtmanspacherGraben07}.\footnote{
    I.e., (1) stationarity: KMS states do not change in time; (2) ergodicity: KMS states are stable under parametric perturbations; (3) mixing: KMS states do not memorize temporal correlations.
}
Interestingly, these equilibrium states can be also obtained from the maximum entropy principle subsumed in the Second Law of thermodynamics. Using KMS states as reference states for the construction of contextual observables, one obtains classical quantities such as temperature through contextual emergence \citep{Takesaki70, Primas98, BishopAtmanspacher06}.

In dynamical systems theory, thermal equilibrium (KMS) states find their counterparts as Sinai-Ruelle-Bowen (SRB) states on chaotic attractors \citep{AtmanspacherGraben07, GuckenheimerHolmes83}. In a coarse-grained description, these can be suitably approximated through Markov processes  \citep{AtmanspacherGraben07, vKampen92} implementing the `molecular chaos assumption' as a stability condition.\footnote{
    In fact, the important Boltzmann equation of physical kinetics is essentially a master equation for a Markov process
    \citep{vKampen92, PitaevskiiLifshitz81}.
}

%--------------------------------------- Section -------------------------------------------------------------

\section{Contextual Emergence of Intentionality}
\label{ce}

\begin{figure}[H]
  \centering
  \includegraphics[width=\textwidth]{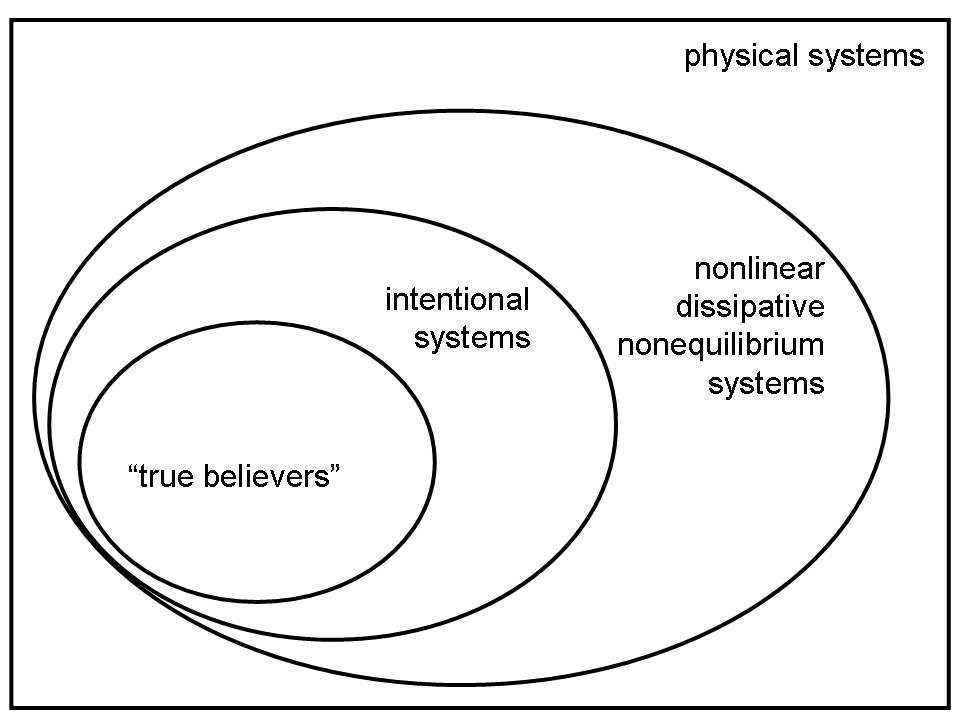}
  \caption{\label{fig:intenthier} Intentional hierarchy.
  }
\end{figure}

Now I have collected all required instruments for the reconstruction of intentionality in terms of contextual emergence. In order to do so, I firstly define the different levels of description. As already pointed out, only two levels do not suffice for that aim. Therefore, I propose the \emph{intentional hierarchy} in \Fig{fig:intenthier} (cf. \citet{JordanGhin06}).

First of all, intentional systems must be physically realized, e.g. as physical symbol systems \citep{Newell80}, presenting the universe of discourse in \Fig{fig:intenthier}. A subset of these are particular dynamical systems, namely nonlinear dissipative nonequilibrium systems. These contain the systems for which Dennett's intentional stance works, i.e. the class of intentional systems. Finally, a subclass of intentional systems are `true believers', systems which really have beliefs and desires in comparison to systems `we may find it handy to treat \emph{as if} they had beliefs and desires' \citep[p. 22]{Dennett89b}.

The subset relations in \Fig{fig:intenthier} also denote entailment relations: being a true believer entails being an intentional system; being an intentional system entails being a nonlinear dissipative nonequilibrium system; and being a nonlinear dissipative nonequilibrium system entails being a physical system. Actually this indicates the necessary relations for contextual emergence: being a physical system is necessary for being a nonlinear dissipative nonequilibrium system; being a nonlinear dissipative nonequilibrium system is necessary for being an intentional system; and being an intentional system is necessary for being a true believer.

First of all, a necessary condition for a nonlinear dissipative nonequilibrium system is certainly that such a system is a physical system. Secondly, one  necessary condition for an intentional system is to have a nonequilibrium dynamical system. This is nicely illustrated by means of Dennett's lectern example:

\begin{quote}
For instance, it seems the lectern in this lecture room can be construed as an intentional system, fully rational, believing that it is currently located at the center of the civilized world (as some of you may also think), and desiring above all else to remain at that center.  \citep[p. 23]{Dennett89b}
\end{quote}

Applying the intentional stance in that way to predict the `behavior' of a lectern has not much predictive power as there is no interesting behavior at all. However, also applying the intentional stance, e.g., to an oscillating pendulum fails although the system is actually exhibiting some temporal dynamics. The reason for this is that even a simple oscillation is (in the ideal case without any friction) a stationary and hence an equilibrium state. The only `desire' that can be ascribed to an equilibrium system is `I wish to remain in my current state', which is obviously lacking much predictive power. Relevant attributions of desires and consequently of beliefs therefore require nonequilibrium dynamical systems that attempt to reach some `desired' attractor or steady state.

Another necessary condition for an intentional system is nonlinearity. This becomes evident in the light of the fact that linear dynamical systems exhibit rather trivial behaviors, namely essentially oscillation and relaxation processes. Yet, in order to talk about an intentional system, it must be an individual object, distinguishable from its environment as a dynamically organized complex `pattern of reality' \citep[p. 245]{Primas90a}. Such patterns, like the magnetic surface swimmers from the introduction, can only emerge via processes of self-organization or self-assembly, requiring nonlinear dynamics.

The last necessary condition for an intentional system is dissipation which means that the system can be described through the design stance in a functional way: it maintains its existence by dissipating energy \citep{TschacherHaken07}. Thus, an intentional system must be nonlinear dissipative nonequilibrium systems as a necessary condition.

Finally, the necessary condition for a true believer is to be an intentional system, i.e. a system that is reliably predictable by means of the intentional strategy \citep[p. 29]{Dennett89b}.

To summarize the necessary conditions of the intentional hierarchy: (1) true believers, among other intentional systems must be physical systems which provide the common ground of my argument. From these one has to exclude (2) all equilibrium systems as they do not exhibit any temporal behavior that could lead to prediction problems. (3) A nonequilibrium system must be governed by nonlinear physical laws that allow for structure formation through self-organization and self-assembly; without nonlinearity one cannot speak about an individual system with complex temporal behavior. (4) Nonlinear nonequilibrium systems must be dissipative in order to be describable by the design stance: they are individual patterns functioning as energy-dissipators. (5) Intentional systems are systems for which the intentional strategy works. (6) True believers are necessarily intentional systems. The following subsections elucidate how to implement stability criteria as sufficient conditions for the contextual emergence of intentionality.

%--------------------------------------- Section -------------------------------------------------------------

\subsection{Contextual emergence of phenomenal laws}
\label{ce1}

Thermal equilibrium systems as discussed in the previous section are stationary systems, i.e.  their statistical states do not change in time. However, for the discussion of intentionality at the level of physical systems one has to look at time-dependent solutions of the evolution law for statistical states.\footnote{
    This law is the Boltzmann equation, a master equation for the probability distribution density of molecules in phase space. In the context of thermal equilibrium the stationary solution of the Boltzmann equation is the ideal gas law \citep{PitaevskiiLifshitz81}.
}

Such solutions are not available in the context of thermal equilibrium which has therefore to be relaxed toward \emph{local equilibrium} holding within the fluid's volume elements. Under this assumption, a nonlinear differential equation describing the dynamics of an ideal fluid, published by Leonhard Euler in 1757, is contextually emergent.\footnote{
    Time-dependency of the Boltzmann equation is described by the so-called substantial derivative consisting of a time derivative and the product of the fluid's velocity and a spatial gradient. For local equilibrium, the nonlinear Euler equation emerges after plugging in the fluid's mean velocity to the substantial derivative which creates a mathematical self-reference \citep{PitaevskiiLifshitz81}. Similar examples of emergent nonlinearity through self-reference have been discussed by \citet{Primas90b}.
}

Yet, the hydrodynamical Euler equation does not account for friction and dissipation.\footnote{
    This became manifest whence Euler who was appointed to the Royal Prussian Academy of Science in 1741 by king Frederick II, was asked for a scientific expertise about the fountains in Potsdam Sanssouci gardens. He calculated the dimensions of the required wooden pipes using his hydrodynamic equations. After building pipes and fountains in way he had suggested they all cracked. This was because Euler neglected friction and dissipation. \citep{Gotz83}
}
In order to achieve this, one has to choose another context where these details are not longer considered to be irrelevant.\footnote{
    Formally, this is carried out by a series expansion of the statistical states around local equilibrium \citep{PitaevskiiLifshitz81}.
}
The resulting evolution law is the so-called Navier-Stokes equation for the velocity field, which is, like the Euler equation, a nonlinear phenomenal law, describing convection and dissipation in fluids. Contextual properties that are due to the Navier-Stokes equation are particularly fluidity, surface tension, convection and waves. For a more general discussion of the contextual emergence of fluid dynamics see \citet{Bishop08, Bishop12}.

%--------------------------------------- Section -------------------------------------------------------------

\subsection{Contextual emergence of complex systems}
\label{ce2}

The contextually emergent, phenomenal nonlinear dissipative Navier-Stokes equation is the core of the self-assembly model of \citet{BelkinGlatzEA10} for magnetic surface swimmers \citep{SnezhkoAransonKwok06, BelkinSnezhkoEA07, SnezhkoBelkinEA09}. Together with the conservation of fluid\footnote{
    Accounted for by a continuity equation.
}
and with Newton's laws for linear acceleration and torque, coupled to the alternating magnetic forcing, these physical laws describe the processes of self-organization and self-assembly of magnetic surface swimmers as individual complex patterns that can be subjected to the intentional strategy.

\begin{figure}[H]
\centering
\subfigure[]{\includegraphics[width=0.2\textwidth]{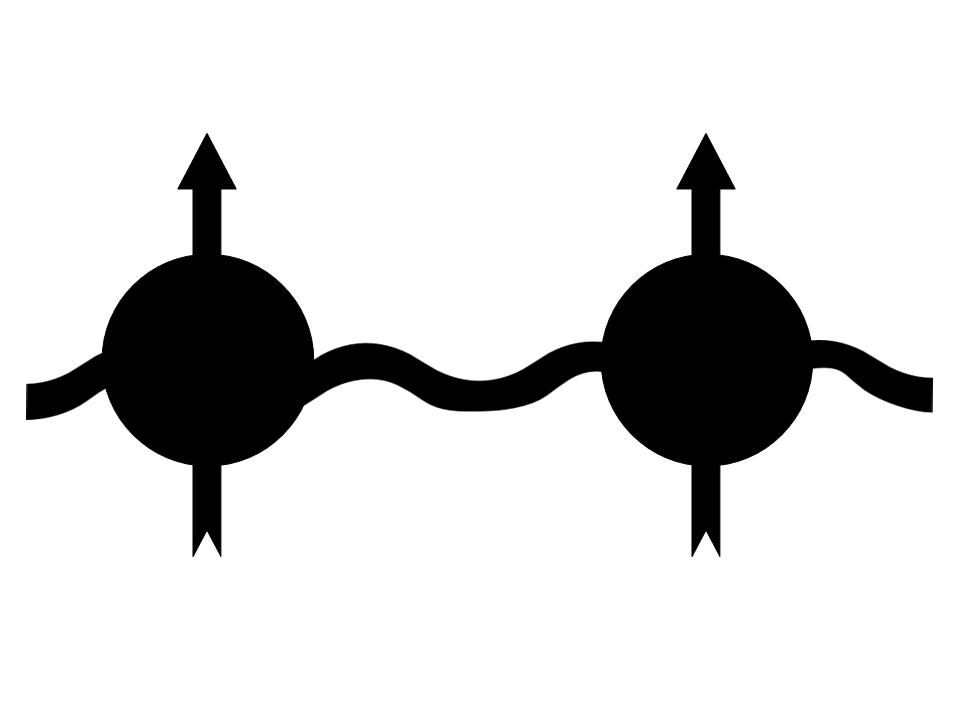}} \hspace{0.2\textwidth}
\subfigure[]{\includegraphics[width=0.2\textwidth]{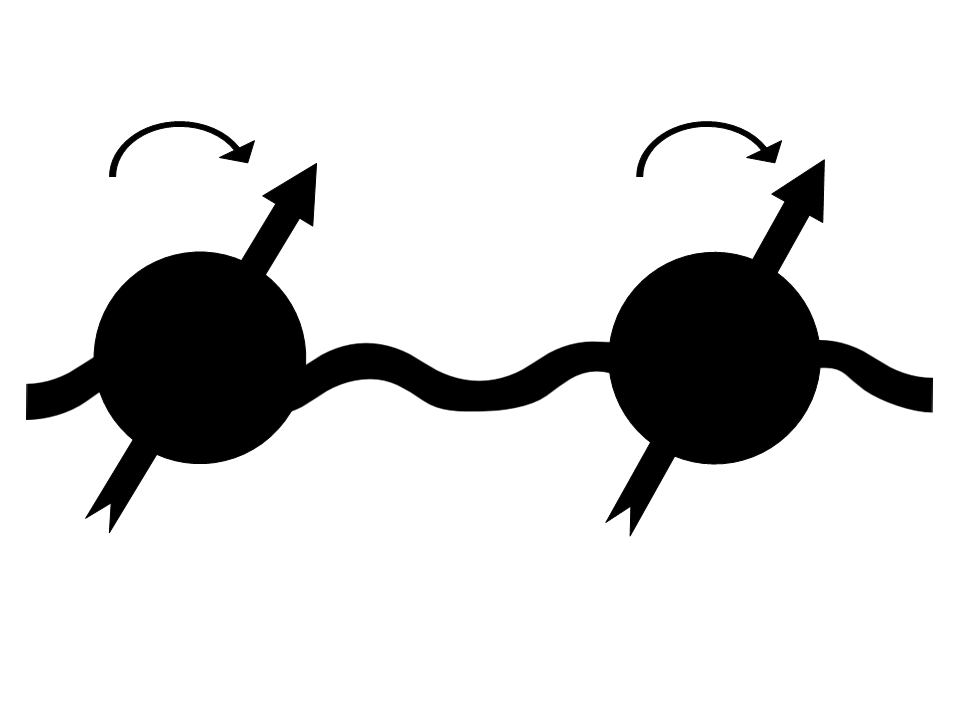}} \\
\subfigure[]{\includegraphics[width=0.2\textwidth]{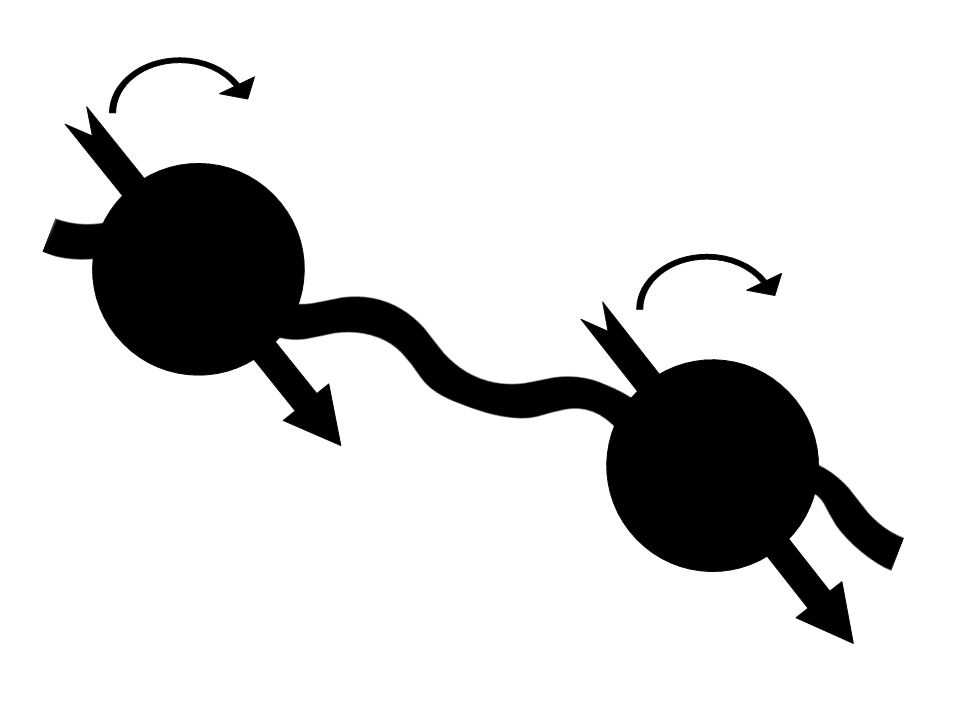}} \hspace{0.2\textwidth}
\subfigure[]{\includegraphics[width=0.2\textwidth]{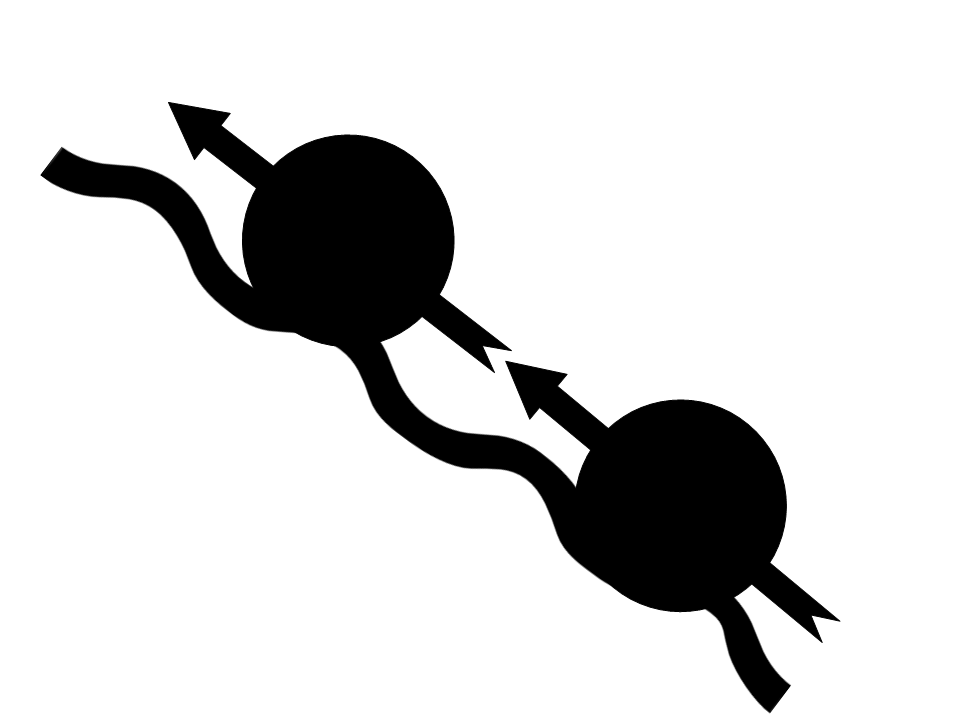}}
  \caption{\label{fig:chain} Self-assembly of magnetic surface swimmer's chain structures. (a) Alignment of nickel spheres' magnetic moments along a static external magnetic field. (b) Rotation of spheres' magnetic moment induced by a periodic external magnetic field. (c) Pumping of surrounding water caused by rotating spheres. (d) Alignment of spheres' magnetic moments to form a dipole chain.
  }
\end{figure}

\begin{figure}[H]
  \centering
  \includegraphics[width=0.6\textwidth]{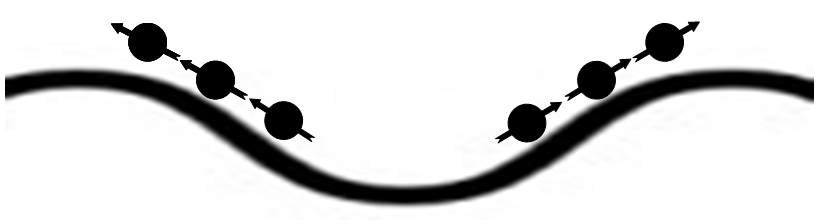}
  \caption{\label{fig:segment} Self-assembly of magnetic surface swimmer's segment structures. Two anti-ferromagnetic segments drag and pull the surrounding water under the external periodic driving.
  }
\end{figure}

These pattern formation processes require a hierarchy of self-assembly steps that are depicted in Figs. \ref{fig:chain} and \ref{fig:segment}.

Figure \ref{fig:chain}(a) shows how ferromagnetic microspheres that are supported by the fluid's --- contextually emergent ---- surface tension align along a static external magnetic field. The repelling forces prevent them from approaching each other too close. However, for a periodically modulated magnetic field the situation is different. According to \Fig{fig:chain}(b) the magnetic forcing causes the spheres to rotate. Thereby, a velocity gradient is created in the surrounding fluid. Due to the presence of --- contextually emergent ---- friction this gradient becomes amplified until it leads to relocation of flowing fluid as shown in \Fig{fig:chain}(c). At some point in time, the magnetic moments of two spheres swimming at different surface heights become attractive. Then they align as displayed in \Fig{fig:chain}(d). This process continues for all spheres at the same side of the surface scarp until they form a magnetic chain.

Through their rotations the spheres exert pressures on the water surface yielding a wave-like pattern that acts as an \emph{order parameter} in synergetics \citep{Haken83a}: due to gravitational force, chains become accelerated along the wave scarps such that they align in parallel along one wave front. Accordingly, wave crests get depleted from microspheres by this interaction between the surface shape and the particles. This leads to the formation of segments as shown in \Fig{fig:segment}.\footnote{
    Where two succeeding segments are anti-ferromagnetically oriented.
}

\footnotemark
    \footnotetext{Cf. \texttt{http://mti.msd.anl.gov/highlights/snakes/coupling.gif}  }

Every two adjacent segments in \Fig{fig:segment} act like a pair of bellows operated by the periodic magnetic forcing that pumps water along the main axis of the surface swimmer. Therefore, the whole structure acts like two linear motors creating a quadrupolar vortex field. As long as both tails produce the same stream, the magnetic surface swimmer stalls somewhere swirling the surrounding fluid. However, for a critical frequency of the driving field this symmetry is spontaneously broken and the object begins moving around. Moreover, this symmetry breaking has also been caused by bringing a plastic bead into the swimmer's environment. At some point in time this bead becomes attached to the head of a structure that transforms into a moving and hunting predator.

So far all these self-assembly processes can be conceived as the necessary conditions for the emergence of a magnetic surface swimmer. The contingent contexts and their stability criteria as sufficient conditions are rather concealed. Here, the context is given by the particular experimental setup imposing several constraints upon the system and its theoretical investigation \citep{Bishop08, Bishop12}. Surface swimmers would not emerge in a random combination of nickel particles, water and a magnetic field. One important contextual constraint is the amount of water in the beaker which must not be too deep.\footnote{
    This constraint is expressed by the so-called shallow water approximation for solving the Navier-Stokes equation \citep{BelkinGlatzEA10}.
}

Though, an even more crucial point is that magnetic surface swimmers in particular and intentional systems in general exhibit characteristic contextual symmetries. These can be geometric symmetries such as the global axial symmetry from head to tail or the local translational symmetries between adjacent segments or between adjacent chains within one segment of a surface swimmer. These can also be material symmetries such as the permutation symmetry of constituents: One could easily exchange the positions of two nickel spheres $A$ and $B$ or of two water molecules $C$ and $D$ without changing the entire structure. Or likewise one could exchange one of these building blocks with any other one taken from the universe without perturbing the systems' functionality. The existence of symmetries indicates that intentional systems are functional patterns which can be described by means of the design stance. Therefore, these systems belong to an ontologically higher level of description than the underlying physical processes of self-organization and self-assembly. Patterns are equivalence classes of physical entities that can be multiple realized (in the sense of supervenience). Symmetry transformations leave the magnetic surface swimmer invariant because they are irrelevant for the functioning and predictability of the structures.

This accords precisely with Dennett's objection to the Martian argument that there is  `something perfectly objective: the \emph{patterns} in human behavior' \citet[p. 25]{Dennett89b}. However, this objectivity is induced by the invariance of structures against material permutations which is contextually grounded in the observer's pattern matching capabilities \citep{Primas90a, Primas98}\footnote{
    Consequently, there are no `Laplacean methods' for predicting complex systems at the physical stance because physical science is constituted by certain regulative principles describing how observers deal with experimental arrangements and with phenomenal laws under perspective decisions \citep{Primas90a}. Also imagined Martian `super-physicists' had to develop statistical physics and hydrodynamics in order to manage physics.
}
and the observer's perspective evaluation of relevant properties.\footnote{
    Cf. \citet[p. 18]{Dennett89b}.
}

%--------------------------------------- Section -------------------------------------------------------------

\subsection{Contextual emergence of intentional systems}
\label{ce3}

As another example for a nonlinear dissipative nonequilibrium system one may next consider a damped pendulum that approaches its resting state in the course of damped oscillations. Would an observer predict this rather trivial behavior by means of the intentional stance? Probably yes, at least at a first glance, saying that the system desires to come to rest, believing that damped oscillations are the most rational pathway toward this goal. But after oscillations have died out, the system becomes similarly uninteresting as the lectern in the center of the universe. What makes the difference to the much more appealing magnetic surface swimmers?

The obvious difference is that the surface swimmers present an open system that is constantly pumped with magnetic energy. This energy has to be dissipated by the generation of heat. In this respect the surface swimmers resemble another hydrodynamic system, namely Rayleigh-B\'enard convection \citep{Bishop08, TschacherHaken07}. Here, a fluid layer is heated from the bottom and cooled at the top, thereby prescribing a temperature gradient as a \emph{control parameter}. When this gradient is below a critical threshold the system clears this perturbation through heat conduction. However, when the temperature gradient exceeds this threshold, the system undergoes a dynamic instability where convection sets in. Depending on the constraints of the experimental setup, different modes of convection compete against each other until a winning mode shapes up as an \emph{order parameter} enslaving the inferior modes \citep{Haken83a, TschacherHaken07}.

\Citet{TschacherHaken07} use this paradigmatic example of synergetics for illustrating their concept of \emph{apparent intentionality}:\footnote{
    See also the neural network account of \citet{HakenTschacher10}.
}

\begin{quote}
Heat convection is realized through the extended coordinated motions --- so-called roll patterns --- of the components in the fluid system \dots self-organization in this system is caused by environmental conditions --- in this case, by the difference in temperature. In turn, however, self-organization exerts an influence on the environment as well. Because the self-organizing system tends to decrease the temperature difference, the environment must react in order to maintain the prescribed difference, i.e. the prescribed control parameter. Hence, the emergence of order parameters may be seen as having a virtual impact on the control parameters which led to order in the first place. In addition to the circular relation between order parameters and components as specified in synergetics, a second circularity is found in systems that thrive on gradients \dots. \Citep[p. 6]{TschacherHaken07}
\end{quote}

In other words, the control parameter, namely the temperature gradient, causes the self-organization of the order parameter, i.e. convection roll patterns. The order parameter in turn attempts to clear the gradient by means of dissipation. This circular causality (or downward causation \citep{Bishop08, Bishop12}) strongly resembles the intentional strategy where beliefs become manifest through the desires attributed to a system, whereas desires become manifest through the beliefs attributed to a system \citep[p. 8]{Dennett78a}.

Here are another two necessary ingredients for the contextual emergence of intentional systems: the patterns (e.g. convection rolls in the Rayleigh-B\'enard system or magnetic surface swimmers) must respond to externally applied gradients, i.e. the system must be open, forced by the environment. And in addition, the system must be in a state of criticality where it has different `choices' how to clear the exerted gradients.

The latter point gives rise to the sufficient condition and its implementation as a stability criterion, which is nothing else than the Second Law of thermodynamics, restated and extended to nonequilibrium dynamical systems:

\begin{quote}
The thermodynamic principle which governs the behaviour of systems is that, as they are moved away from equilibrium they will utilize all avenues available to counter the applied gradients. As the applied gradients increase, so does the system's ability to oppose further movement from equilibrium. \citep[p. 29]{SchneiderKay94}
\end{quote}

This means that a system expelled from equilibrium by external gradients responds in the \emph{most optimal way} to get rid of the gradients. \Citet[p. 7]{TschacherHaken07} write:

\begin{quote}
\begin{itemize}
  \item \emph{Self-organisation and thermodynamics}: Self-organization phenomena can be treated in connection with a restated second law of thermodynamics. The consequence of spontaneous increase of order in an open system can be discussed as a corollary of the second law.
  \item \emph{The functionalist view of self-organization}: The restated second law regards pattern formation in the service of gradient reduction. Thus, pattern formation is functional; it may `look' intentional.
  \item \emph{Optimality of self-organization}: In principle, there may be several patterns, each of which can be functional in reducing the gradients imposed on the system. These patterns can be attributed to different efficiencies. If a specific pattern provides the optimal (most efficient) way to dissipate the gradients, precisely this pattern will be realized in the system. The better dissipative pathway is the preferred one.
\end{itemize}
\end{quote}

Therefore, the sufficient condition for the contextual emergence of intentional systems is precisely Dennett's rationality assumption:

\begin{quote}
A prediction relying on the assumption of the system's rationality is relative to a number of things. First, rationality here so far means nothing more than optimal design relative to a goal or optimal weighted hierarchy of goals \dots and a set of constraints \dots \citep[pp. 4f]{Dennett78a}
\end{quote}

In this sense, the Rayleigh-B\'enard system behaves rational since it `prefers' convection rolls over heat conduction as the optimal pathway for dissipating energy. In the same way, the magnetic surface swimmers behave rational as they `prefer' swimming around than standing still. For them, swimming around and hunting nickel spheres is even more rational because they could grow, thus becoming faster and leveraging their efficacy for dissipating energy.\footnote{
    Another criterion that has been suggested for the emergence of intentionality is self-sustainment \citep{JordanGhin06, JordanHeidenreich10}: Self-sustaining systems have the capability to retroact upon their defining contingent contexts. This might introduce another level within the intentional hierarchy \Fig{fig:intenthier}.
}

%--------------------------------------- Section -------------------------------------------------------------

\subsection{Contextual emergence of true believers}
\label{ce4}

Intentional systems such as the Rayleigh-B\'enard system or magnetic surface swimmers can be treated by the intentional stance \emph{as if} they had beliefs and desires. How to distinguish them from systems that \emph{really} have beliefs and desires \citep[p. 22]{Dennett89b}, i.e. from `true believers' \citep[p. 19]{Dennett89b}? This must be achieved by remedying the irritating observer-dependence of intentionality \citep[pp. 23f]{Dennett89b}, or in other words by making intentional ascriptions objective. But what does \emph{objectivity} mean in contemporary, operationalistic science?

Basically, objectivity refers to invariance through symmetry: a natural law holds objectively because it holds always and everywhere in the universe and for every observer who is able and competent to replicate an observational setup. In other words, objective judgments are invariant under exchanging observers.

This must also hold for the attribution of intentional states as well in order to talk about true believers. Therefore, a true believer is an intentional system ---  providing the necessary condition ---  which is stable under permutation of observers --- implementing the sufficient condition as a stability criterion. In particular, a true believer must be able to ascribe intentionality to its own behavior (reflexivity) and when an observer $A$ adopts the intentional stance for predicting $B$'s behavior, $B$ must be able to adopt the intentional stance for the prediction of $A$'s behavior as well (symmetry). Finally when $A$ ascribes intentional states to $B$ and $B$ does the same with $C$, $A$ must be able to attribute intentionality to $C$ also (transitivity). Accordingly, true believers comprise an equivalence class through `\emph{the unavoidability of the intentional stance with regard to oneself and one's fellow intelligent beings}' \citep[p. 27]{Dennett89b}. They reciprocally ascribe intentional states to each other in the course of reliable and respectful communication.

%--------------------------------------- Section -------------------------------------------------------------

\section{Discussion}
\label{disc}

Illustrated by means of the intriguing example of self-assembling magnetic surface swimmers \citep{SnezhkoAransonKwok06, BelkinSnezhkoEA07, SnezhkoBelkinEA09} which can be described --- at least to some extent --- by attributing beliefs and desires to them according to the intentional strategy \citep{Dennett78a, Dennett89b}, I have reconstructed a hierarchy of necessary and sufficient conditions for the applicability of this strategy (\Fig{fig:intenthier}). It has turned out that these conditions demand a similar operationalistic treatment as Dennett's approach. While the intentional strategy is adopted by an observer in order to predict the behavior of a system under consideration, the different levels of the intentional hierarchy emerge by selecting appropriate contexts through an observer. These contexts define the relevant conditions that are sufficient for each level by imposing stability constraints upon the subjacent level in the framework of contextual emergence \citep{BishopAtmanspacher06, AtmanspacherBishop07}.

At the lowest level of the hierarchy, phenomenal physical laws emerge for the coarse-grained description of open, nonlinear, and dissipative nonequilibrium systems in critical states. One level higher, dynamic patterns, such as, e.g., magnetic surface swimmers, are contextually emergent as they are invariant (and hence stable) under certain symmetry operations. Again one level up, these patterns select optimal pathways for the dissipation of energy that is delivered by external gradients. Thus, these patterns behave rationally in accordance with the restated Second Law of thermodynamics as stability criterion. At the highest level, true believers are intentional systems that are stable under exchanging their observation conditions.

All these different levels of description are pervaded by one thread: the prevalence of the \emph{observer}. According to Dennett it is always an observer who employs one of the three stances: the physical stance, the design stance, or the intentional stance, e.g. facing a chess computer:

\begin{quote}
The decision to adopt the strategy is pragmatic, and is not intrinsically right or wrong. One can always refuse to adopt the intentional stance toward the computer, and accept its checkmates. One can switch stances at will without involving oneself in any inconsistencies or inhumanities, adopting the intentional stance in one's role as opponent, the design stance in one's role as redesigner, and the physical stance in one's role as repairman.
\citep[p. 7]{Dennett78a}
\end{quote}

Likewise, it is always the observer who selects a suitable context for describing properties of natural phenomena. And finally, only observers are able to treat each others as intentional systems. Thus I conclude that \emph{true believers are real observers} \citep{Graben2011}.

%--------------------------------------- Section -------------------------------------------------------------

\section{Acknowledgements}

This work has flourished over the last ten years in fruitful discussions with Bettina Stangneth and Stefan Frisch. I am gratefully indebted to Michail Rabinovich and Igor Aranson for drawing my attention toward magnetic surface swimmers. I thank Yasemin Erden for her kind invitation to submit an elaborated version of an AISB 2013 presentation to the \emph{Journal of Consciousness Studies}. For the last four years I acknowledge support by a Heisenberg Fellowship of the German Research Foundation DFG  (GR 3711/1-2).

\nocite{Metzinger00}

% \bibliography{PbG}

\end{document}